\documentclass[12pt,a4paper,twocolumn]{narms}

\usepackage{subfigure} 
\usepackage{epsfig}
\usepackage{timesmt}   
\usepackage{chicaco}

\usepackage{psfrag}

\makeatletter
\renewcommand{\section}{\@startsection%
{section}%
{1}%
{0mm}%
{- \baselineskip}%
{0.15\baselineskip}%
{\normalfont\normalsize}}%

\renewcommand{\subsection}{\@startsection
{subsection}%
{2}%
{0mm}%
{-\baselineskip}%
{0.15\baselineskip}%
{\normalfont\normalsize}}%
\makeatother

\newcommand{\vn}[1]{\mbox{\boldmath$#1$}}
\newcommand{\bsigma}{\vn{\sigma}}
\newcommand{\bvarepsilon}{\vn{\epsilon}}
\newcommand{\bepsilon}{\vn{\varepsilon}}
\newcommand{\bkappa}{\vn{\kappa}}
\newcommand{\tr}{{\rm tr}}
\newcommand{\dev}{{\rm dev}}

\renewcommand{\matrix}[1]{{\bf{#1}}}

\begin{document}

\title{From DEM Simulations towards a Continuum Theory of Granular Matter}
\author{\large {Stefan Luding}\\
{\em Institute for Computer Applications 1, 
     Pfaffenwaldring 27, 70569 Stuttgart, Germany }\\
{    e-mail: lui@ica1.uni-stuttgart.de -- 
     URL: http://www.ica1.uni-stuttgart.de/\~{}lui}
}
\date{}

\abstract{ABSTRACT: 
  One essential question in material sciences is how to bridge
  the gap between the microscopic picture and a macroscopic description.
  The former involves contact forces and deformations, whereas the
  latter concerns tensorial quantities like the stress or the 
  velocity gradient. \\
  A two-dimensional shear-cell filled with disks of different sizes is 
  examined by means of a ``microscopic'' discrete element method (DEM). 
  Applying a consistent averaging formalism, one can obtain scalar- and
  vector-fields as well as classical tensorial macroscopic quantities like 
  fabric, stress or velocity gradient and, in addition, micropolar 
  quantities like curvature or couple-stress.
}


\maketitle
\frenchspacing   


\psfrag{Title}[][]{}
\psfrag{3358}[][][0.75]{0.8083}
\psfrag{3386}[][][0.75]{0.8149}
\psfrag{3401}[][][0.75]{0.8194}

%

\section{INTRODUCTION}

\begin{figure}[htb]
  \begin{center}
    \epsfig{file=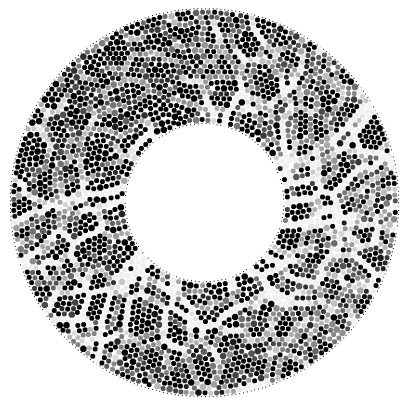,height=5.6cm}
  \end{center}
  \caption{Representative snapshot of the model system.
           Light and dark grey indicate large and small values
           of the contact potential energy, respectively.}
  \label{fig:shearcell}
\end{figure}

The macroscopic balance equations for mass, momentum and energy can
be used for the modeling of the behavior of granular media. However, they
rely on constitutive relations between the physical quantities, in order
to close the system of equations. 
The determination of both the physical quantities, important for 
the behavior of the system, and their inter-relations are subject of current 
research \shortcite{behringer97,herrmann98,vermeer00}.  
One possible way to obtain an observable like, for example, the stress is
to perform experiments with photo-elastic material 
\shortcite{veje99,howell99,howell99b} 
which allows for non-invasive force measurements.
The alternative is to perform discrete particle simulations 
\shortcite{cundall79,herrmann98,latzel00} and to average 
over the ``microscopic'' forces and contact vectors in the simulation.
A typcal snapshot from a two-dimensional DEM simulation is displayed
in Fig.\ \ref{fig:shearcell}.

Besides the trivial definitions for
averages over scalar and vectorial quantities like density, velocity, and
particle-spin, one can find differing definitions for stress
and strain averaging procedures in the literature
\shortcite{bathurst88,kruyt96,liao97b,calvetti97,latzel00,dedecker00}.

The aim of this paper is to review recent results for tensorial,
averaged continuum quantities involving also micro-polar aspects
\shortcite{latzel00b,zervos00}.

In section~\ref{sec:micro-macro} our averaging method is introduced and
applied to obtain some macroscopic fields in section~\ref{sec:scalar}.
Section~\ref{sec:macro} contains the definitions and averaging
strategies for fabric, stress, and elastic deformation gradient.
The particle rotations are taken into account in
section~\ref{sec:rotations}, averages are presented for 
selected macroscopic quantities in section \ref{sec:results}
and material parameters are discussed in section \ref{sec:material_parameters}.

\section{FROM MICRO- TO MACRO-DESCRIPTION}
\label{sec:micro-macro}

In a microscopic, discrete picture, the know\-ledge of the
forces acting on each particle is sufficient to model the dynamics and
the statics of the system. Tensorial quantities like the stress 
$\bsigma$ or the strain $\bepsilon$ are not required for 
a discrete model. 
In order to establish a correspondence to continuum theories 
one has to compute tensorial fields as well as scalar
material properties like, e.\,g., the bulk and shear moduli
\shortcite{cambou95,kruyt96,emeriault97,liao97b,latzel00,luding00d,latzel00b}.
In the following, a consistent averaging strategy is introduced.

\subsection{\em Averaging Strategy}

Most of the measurable quantities in granular materials vary strongly
both in time and on short distances.  At a fixed point in space, one has
either a particle or one finds an empty space.  Therefore, density flips
between the two values zero and $\varrho^p$, where the latter is the material
density.  During the ``homogenization'' presented in the next
subsection, we average over space and time, in order to reduce 
the fluctuations.  The averaging over space is an option in systems
with some translational or rotational invariance (simple shear or ring
shear apparatus); averaging over long times is feasible only in a steady state
situation. 
In a ring shear apparatus, as introduced later in section\ \ref{sec:results}, 
all points at a certain distance from the origin are equivalent to each 
other and, after several shear cycles, the system reaches a quasi-steady 
state.  Averaging over
many snapshots is somehow equivalent to an ensemble average.  However,
we remark that different snapshots are not necessarily independent of
each other \shortcite{latzel00} and the duration of a
simulation might be too short to explore a representative part of the
phase space.

The final assumption for the averaging procedure presented below, is 
that all quantities are smeared out over one particle.  This is a good
approximation for the density, but not for the stress.  However, since 
we aim at macroscopic, averaged quantities, it is not our goal to solve 
for the stress field inside one particle.
Details of the position dependency inside the particles are presumably
smeared out due to the averaging in space and time.

\subsection{\em Averaging Formalism}

The mean value of some quantity $Q$ is defined as
\begin{equation}
Q = \frac{1}{V} \sum_{p \in V} w_V^p V^p Q^p ~,
\label{eq:formalism}
\end{equation}
with the averaging volume $V$ and the particle volume $V^p$. 
$Q^p = \sum_{c=1}^{{C}^p} Q^c$ is a quantity attributed to particle $p$, 
where the quantity $Q^c$ is attributed to contact $c$ of particle $p$ with
${C}^p$ contacts.  The weight $w_V^p$ accounts for the particle's 
contribution to the
average, and corresponds to the fraction of the particle volume that
is covered by the averaging volume.  Since an exact calculation of the
area of a circular particle that lies in an arbitrary volume is rather
complicated, we assume that the boundaries of $V$ are locally
straight, i.\,e.\ we cut the particle in slices,
see~\shortcite{latzel00} for details.
In the 
following section possible choices for observables are summarized 
and discussed.

\section{MACROSCOPIC FIELDS}
\label{sec:scalar}

In the following, the averaging formalism in Eq.\ (\ref{eq:formalism})
is applied to obtain various macroscopic quantities. For example,
the quantity $Q^p$ can be chosen as $\varrho^p$ in order to obtain the
density, as $\varrho^p \vn v^p$ in order to obtain the momentum density,
or as $(1/2) \varrho^p (\vn v^p)^2$ for the kinetic energy density.  
In table~\ref{tab:quantities},
the macroscopic fields as well as particle-attributed quantities 
are collected.
In this study, two tensors with no symbol in between mean a dyadic product,
whereas `$\cdot$' is used for the scalar product, i.\,e.\ the order-reduction 
by one for each of the two tensors at left and right. 
\begin{table}[htb]
  \caption{Macroscopic fields computed using the averaging formalism
           in Eq.\ (\protect\ref{eq:formalism}) using particle properties}
  \label{tab:quantities}
  \begin{center}
    \begin{tabular}{|l|c|}
\hline
macroscopic quantity & $Q^p$ \\ 
\hline
\hline
  volume fraction $\nu$ &
  $1$ \\
  density $\varrho$ &
  $\varrho^p$ \\
  flux density $\nu \vn{v}$ &
  $\vn{v}^p$ \\
  momentum density $\varrho \vn{v}$ &
  $\varrho^p \vn{v}^p$ \\
  kinetic energy density $\frac{1}{2} \varrho \vn{v}^2$ &
  $\frac{1}{2} \varrho^p (\vn{v}^p)^2$ \\
  dynamic stress tensor $^d\matrix{\bsigma}$ &
  $\displaystyle \varrho^p \vn{v}^{p} \vn{v}^{p}$ \\
\hline
    \end{tabular}
  \end{center}
\end{table}
~\vspace{-1.2cm}\\

\subsection{\em Volume Fraction}

As a first example for an averaged scalar quantity, the local volume
fraction $\nu$, see Tab.~\ref{tab:quantities}, is
directly related to the local density via
$
   \varrho \approx \varrho^p \nu \,.
$

\subsection{\em Flux Density}
\label{sec:massflux}

As a tensorial quantity of first order (vector), the velocity field $\vn v$
is obtained when dividing the flux density $\nu \vn v$ by the volume
fraction $\nu$. The momentum density is thus 
$
   \varrho \vn{v} \approx \varrho^p \nu \vn{v}~.
$

\subsection{\em Velocity Gradient}
\label{sec:vgrad}

Given the velocity field, it is possible to compute the velocity
gradient $\vn{\nabla} \vn{v}$, a tensor of second order, by means of 
numerical differentiation. With $\vn{\nabla}\vn{v}$ one has the
components of the symmetric deformation rate 
\begin{equation}
D_{\alpha \beta} =
\frac{1}{2} \left[
              \frac{\partial v_\alpha}{\partial \beta}
             +\frac{\partial v_\beta}{\partial \alpha}
            \right]
\label{eq:Dab}
\end{equation}
and the anti-symmetric continuum rotation rate
\begin{equation}
W_{\alpha \beta} =
\frac{1}{2} \left[
              \frac{\partial v_\alpha}{\partial \beta}
             -\frac{\partial v_\beta}{\partial \alpha}
            \right] ~,
\label{eq:Wab}
\end{equation} 
with the coordinates $\alpha$ and $\beta$.
In the special case of cylindrical symmetry with radial direction $r$
and angular direction $\phi$, Eqs.\ (\ref{eq:Dab}) and (\ref{eq:Wab})
reduce to 
\begin{equation}
D_{r\phi}=\frac{1}{2}
          \left[
              \frac{\partial v_\phi}{\partial r}-\frac{v_\phi}{r}
          \right]
\end{equation}
and 
\begin{equation}
W_{r\phi}=\frac{1}{2}
          \left[ 
              \frac{\partial v_\phi}{\partial r}+\frac{v_\phi}{r}
          \right] ~,
\end{equation} 
see~\shortcite{luding00d} for
details and \shortcite{zervos00} for a similar approach.

\subsection{\em Energy Density}

The kinetic energy density is only a function of the velocity of
the particles in the averaging volume and it is related to the 
trace of the dynamic stress tensor. However, we do not discuss this
quantity here, since the system examined later is quasi-static
so that the dynamic contributions are of minor importance
\shortcite{luding00d,latzel00b}.

\section{MACROSCOPIC TENSORIAL QUANTITIES}
\label{sec:macro}

In this section, the averaged, macroscopic tensorial quantities in our
model system are introduced. The fabric tensor describes the statistics 
of the contact directions, the stress tensor describes the stress 
distribution due to the contact forces, and the elastic deformation 
gradient is a measure for the elastic, reversible deformations due to
the stress.

\begin{table}[htb]
  \caption{Macroscopic tensorial quantities computed with 
           Eq.\ (\protect\ref{eq:formalism}) using 
           contact-attributed properties pre-averaged over
           single particles}
  \label{tab:quant}
  \begin{center}
    \begin{tabular}{|l|c|}
\hline
macroscopic quantity & $Q^p$ \\ 
\hline
\hline
  fabric tensor $\matrix{F}$ &
  $\displaystyle \sum_{c=1}^{{C}^p} \vn{n}^{c} \vn{n}^{c}$ \\
[2mm]

  static stress tensor $\matrix{\bsigma}$ &
  $\displaystyle \frac{1}{V^p} \sum_{c=1}^{{ C}^p} \vn{f}^{c} \vn{l}^{pc}$ \\
[2mm]


  deformation gradient $\matrix{\bvarepsilon}$ ~~ &
  $\displaystyle \frac{\pi h}{V^p} \sum_{c=1}^{{ C}^p} \vn{\Delta}^{pc}
  \vn{l}^{pc} \cdot \matrix{F}^{-1}$ \\
[2mm]
\hline
    \end{tabular}
  \end{center}
\end{table}
~\vspace{-1.4cm}\\

\subsection{\em Fabric Tensor}

In assemblies of grains, the forces are transmitted from one particle
to the next only at the contacts of the particles.
Therefore, the local geometry and direction of each
contact is important \shortcite{latzel00}.
The fabric tensor, in our definition, involves the 
contact normal vector $\vn{n}^c$, 
related to the so-called branch vector via
$\vn{l}^{pc} = a_p \vn{n}^c$, 
with particle radius $a_p$. 
For each contact the dyadic product $\vn{n}^c \vn{n}^c$ is 
used, i.\,e.~a degenerate tensor of order two with non-zero
value in $\vn{n}^c$-direction only.  In average over many 
contacts (and over many particles) one obtains the fabric tensor.

The fabric tensor in table~\ref{tab:quant} is symmetric by
definition and thus consists of up to three independent scalar
quantities in two dimensions. The first of them, the trace (or
volumetric part) $F_V = \tr(\matrix{F}) = (F_{\rm max}+F_{\rm min})$,
is the contact number density, with the major and the minor
eigenvalues $F_{\rm max}$ and $F_{\rm min}$, respectively.  With other
words, one obtains the relation $\tr(\matrix{F})=\nu{ C}$ with
reasonable accuracy for monodisperse particles, where ${C}$ is the 
average number of contacts per particle.
For polydisperse size distribution functions, a multiplicative
correction factor was recently proposed \shortcite{madadi01}, which
depends on the first three moments of the size distribution function only.


The second scalar, the deviator $F_D = F_{\rm max}-F_{\rm min}$, 
accounts for the anisotropy of the contact network to first order,
and the third, the angle $\phi_F$, gives the orientation of 
the ``major eigenvector'', i.\,e.\ the eigenvector corresponding to 
$F_{\rm max}$, defined with respect to the radial outwards direction
for cylindrical symmetry.  In other words, the contact probability 
distribution is proportional to the function $F_V + F_D \cos(2(\phi-\phi_F))$
\shortcite{latzel00}, when averaged
over many particles. Note that this approximation, a fourier expansion 
up to second order, is not always sufficient, i.\,e. a more general
fabric tensor of rank four should be used \shortcite{mehrabadi88,herrmann98}. 


\subsection{\em Stress Tensor}
\label{sec:stress}

The stress tensor $\bsigma$ is a measure for the force 
$\vn{f} = \bsigma \cdot \vn{n}$, acting on an imaginary 
surface with normal $\vn{n}$, per unit area.  A force is
equivalent to some momentum transfer per unit time, so that 
also the momentum flux density contributes to the stress
in a dynamic way.  The dynamic stress, see table\ \ref{tab:quantities},
corresponds to the stress in an ideal gas, whereas the static stress,
see table\ \ref{tab:quant}, is caused by the transfer of forces from 
a contact to the center of mass of a particle.

In order to account for both the stress and the distance
of transfer, the static component of the stress tensor 
\shortcite{rothenburg81,satake88,kruyt96,latzel00}
is defined as the dyadic product of the force $\vn{f}^{c}$ acting 
at contact $c$ with the corresponding branch vector, see table
\ref{tab:quant}.
Note that the formerly introduced surface quantity is here
expressed as the sum over all particles in the averaging volume,
with the respective weight factor in Eq.\ (\ref{eq:formalism}),
for a detailed derivation see \shortcite{latzel00}.


\subsection{\em Elastic Deformation Gradient}

In order to reach the final goal, i.\,e.~to measure the material properties 
of a granular ensemble, one is interested, e.\,g.,~in the stress-strain 
relationship of the material.  The
strain $\bepsilon$ can be obtained by time integration of the velocity
gradient, see subsection~\ref{sec:massflux}, and subsequent
symmetrization and linearization. Here, an alternative quantity is
introduced by application of ``Voigt's hypothesis'', i.\,e.\ assuming that 
the deformation is uniform and that every particle displacement conforms
to the corresponding mean displacement field, but fluctuates about it
\shortcite{liao97b,latzel00}.  The deformation gradient in 
table\ \ref{tab:quant} is calculated for the special case of 
two-dimensional disks with height $h$.

This relates the actual deformations to a virtual, stress-free reference state
where all contacts start to form, i.\,e.\ the particles are just touching,
see~\shortcite{latzel00} for a detailed derivation. 
The result is a non-symmetric tensor $\bvarepsilon$, which is {\it not}
the strain. Instead, we refer to it as the elastic deformation
gradient, since it accounts only for reversible (elastic) deformations. 


\section{ROTATIONAL DEGREES OF FREEDOM}
\label{sec:rotations}

Due to the particles' surface roughness, forces are transmitted also
tangentially and thus granular particles will rotate. Therefore,
also micro-polar macroscopic quantities related to the rotational 
degrees of freedom are of interest \shortcite{latzel00b,zervos00}.
The particle rotation, the couple-stress and the curvature are
discussed in this section.

\begin{table}[htb]
  \caption{Tensorial quantities, connected to the rotational degrees
           of freedom, computed with Eq.\ (\protect\ref{eq:formalism}) }
  \label{tab:rot}
  \begin{center}
    \begin{tabular}{|l|c|}
\hline
macroscopic quantity & $Q^p$ \\ 
\hline
\hline
  spin density $\nu \vn{\omega}$&
  $\displaystyle \vn{\omega}^{p}$ \\
[2mm]

  couple stress $\matrix{M}$ &
  $\displaystyle \frac{1}{V^p} \sum_{c=1}^{{C}^p}
                  \left( \vn{l}^{pc}\times\vn{f}^c\right) \vn{l}^{pc}$ \\
[4mm]

  elastic curvature $\bkappa$ &
  $\displaystyle  \frac{\pi h}{V^p} \sum_{c=1}^{C^p}
  ( \vn{l}^{pc} \times \vn{\Delta}^{pc}) \vn{l}^{pc} \cdot \vn{F^{-1}}$\\
[2mm]
\hline
    \end{tabular}
  \end{center}
\end{table}
~\vspace{-1.2cm}\\

\subsection{\em Particle Rotation}

The mean angular velocity of the particles is obtained by
using $Q^p=\vn{\omega}^p$, see table\ \ref{tab:rot}.  Note that
$\vn{\omega}$ also contains the continuum angular velocity  
\begin{equation}
\vn{\omega}_c = \matrix{I} \times \matrix{W} ~,
\label{eq:omegac}
\end{equation}
i.\,e.\ the vector product `$\times$' of the continuum rotation 
rate from Eq.\ (\ref{eq:Wab}) and the unit tensor $\matrix{I}$. 
The excess rotation, or particle eigen rotation, with respect to the
underlying mean motion, is thus
\begin{equation}
\vn{\omega}^*=\vn{\omega} - \vn{\omega}^c ~,
\end{equation}
i.\,e.\ the total mean spin minus the continuum spin.

\subsection{\em Couple stress}

In the framework of a Cosserat continuum
\shortcite{zervos00} one has, 
in addition to the stress, also the couple stress
$\matrix{M}$. When an applied stress leads to a deformation, an 
applied couple-stress causes a rotational motion. 
It can thus be defined in analogy to the stress by
replacing the force by the torque due to the tangential
component of the force, see table\ \ref{tab:rot}.
In a two dimensional system with cylindrical symmetry, only the 
components $M_{zr}$ and $M_{z\phi}$ of the tensor survive. 
Note that $\matrix{M}=\matrix{0}$, when the sum of the torques acting on
spherical particles vanishes by definition in static equilibrium. 

\subsection{\em Elastic Curvature}

In analogy to the elastic deformation gradient, we define the
curvature $\bkappa$ by replacing the overlap with the respective
axial vector, $\vn{l}^{pc} \times \vn{\Delta}^{pc}$, see table\ \ref{tab:rot}.
This leads to a measure for the reversible or ``frozen in'' rotations
in the system.  Consequently, in static equilibrium of spherical particles,
one has $\bkappa=\matrix{0}$.

\section{RESULTS}
\label{sec:methods}
\label{sec:results}


The elementary units of granular materials are mesoscopic grains.
With DEM~\shortcite{latzel00,latzel00b} the grains are treated as rigid
particles but their local deformation at their contact points is
realized as virtual overlaps.
We relate the interaction forces to this overlap $\delta$ and
to the tangential displacement of two particles during contact. The
force laws used are material dependent and have to be validated by
comparison with experimental measurements
\shortcite{foerster94,labous97,falcon98,luding98c}.

\subsection{\em The Model System}

In the simulations presented in this study, a two-dimensio\-nal
Couette shear-cell is used.
$N$ disks are confined between an outer ring and an inner ring
with radius $R_o$ and $R_i$, respectively. The particles are of
slightly different size $d_{\rm small}=7.42$\,mm and 
$d_{\rm large}=8.99$\,mm, in order to reduce ordering effects.  
These boundary conditions are based on an experiment
\shortcite{howell99,veje99}; for more details and other simulations,
see~\shortcite{latzel00,luding00d}.

The outer wall is fixed and the inner wall rotates and thus introduces a
slow shear deformation to the system.  The simulations are started in a dilute
state with an extended outer ring while the inner ring already rotates
counter-clockwise with constant angular velocity $\Omega = 2 \pi / T_i
= 0.1$\,s$^{-1}$ and period $T_i=62.83$\,s. The radius of the outer
ring is reduced within about two seconds to reach its desired value
$R_o$ and thereafter it is kept fixed. 
Averages are performed after about three rotations at $t=180$\,s (to
get rid of the arbitrary initial configuration), and during about one
rotation, until $t=239$\,s.  In Fig.\ \ref{fig:shearcell}, a typical
snapshot is displayed, where the potential energy of the particles 
is coded in greyscale so that the stress-chains become visible.

Different global volume fractions $\bar \nu$ were examined in the 
simulations. Here, we present data from three different simulations
A, B, and C with $\bar \nu=0.8084$, $0.8149$ and $0.8194$,
respectively. For the different simulations the number of large and
small particles is $N_{\rm small}$: $2511$, $2545$, $2555$ and
$N_{\rm large}$: $400$, $394$, $399$.
For the calculation of the global volume fraction, the small particles
glued to the wall are counted with half their volume only, and thus
contribute with $\bar \nu_{\rm wall} = 0.0047$ to $\bar
\nu$. For more details and material parameters, see~\shortcite{latzel00}. 

\subsection{\em Volume Fraction}

\begin{figure}[htb]
  \begin{center}
     \psfrag{Title}{}
     \psfrag{nu}[][]{$\bar{\nu}$}
     \psfrag{xlabel}[][]{$\tilde{r}$}
     \psfrag{ylabel}[][]{$\nu$}
     ~\vspace{-1.1cm}\\
     \epsfig{file=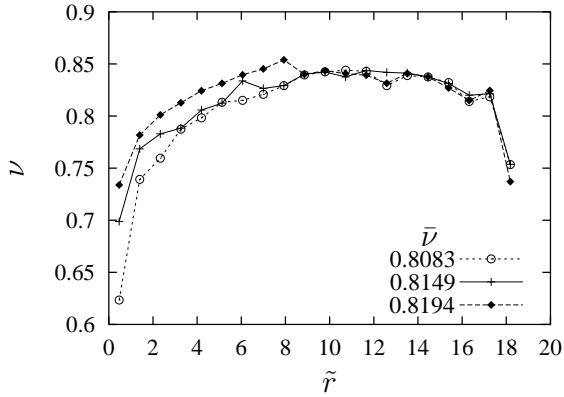,width=6cm,angle=-90,clip=true}
  \end{center}
  \caption{Volume fraction $\nu$, plotted against the dimensionless distance
    from the origin $\tilde{r}=(r-R_i)/\tilde{d}$, with the typical diameter
    $\tilde{d}=8$\,mm, for different initial global densities $\bar{\nu}$}
  \label{fig:vol_frac}
\end{figure}

In Fig.~\ref{fig:vol_frac}, the volume fraction is plotted against the
rescaled distance from the inner ring.  Starting from an initially uniform 
volume fraction, a dilated shear-zone forms near to the inner wheel as a
consequence of the applied shear.  This effect is less pronounced for
higher initial global densities and, in the outer region of the shear-cell
($\tilde{r} > 10$), the structure of the packing remains frozen, i.\,e.\
not much reorganization takes place within the duration of the
simulation.

\subsection{\em Velocity Field}

\begin{figure}[htb]
  \begin{center}
    \psfrag{Title}[][]{}
    \psfrag{ylabel}[][]{$v_{\phi}/\Omega R_i$}
    \psfrag{xlabel}[][]{$\tilde{r}$}
     ~\vspace{-1.0cm}\\
    \epsfig{file=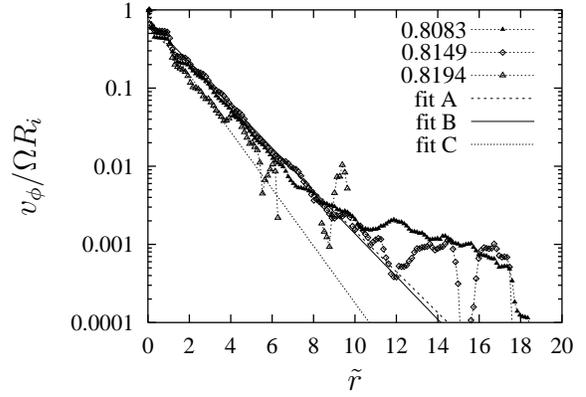,width=6cm,angle=-90,clip=true}
  \end{center}
  \caption{Tangential velocity $v_\phi$ normalized by the
    velocity of the inner wheel $\Omega R_i$, plotted against
    $\tilde{r}$. The lines are the fits to the simulation data for
    $\tilde{r}=0.25 {\rm ~to~} 8.1$, using $ v_\phi(r) = v_0
    \exp(-\tilde{r}/s)$ with $v_{o}$: $0.670$, $0.756$, $0.788$ and
    $s$: $1.662$, $1.584$, $1.191$, for increasing density, as given
    in the inset, respectively }
  \label{fig:dens1}
\end{figure}

In Fig.~\ref{fig:dens1}, the tangential velocity is plotted against the 
rescaled distance from the inner ring.  The simulation data are fitted by
a function $v_\phi(r) = v_0 \exp(-\tilde{r}/s)$ with $v_{o}$:
$0.670$, $0.756$, $0.788$ and $s$: $1.662$, $1.584$, $1.191$, thus
showing an exponential profile corresponding to the shear band.
The shear band, has a width of a few 
particle diameters, before the velocity $v_\phi$ reaches the noise level.

\subsection{\em Fabric Tensor}

\begin{figure}[htb]
~\vspace{-0.8cm}\\
    \psfrag{ylabel}[][][0.70]{$\tr(\matrix{F})$}
    \psfrag{ylabel_b}[][][0.70]{$\dev(\matrix{F})/\tr(\matrix{F})$}
    \psfrag{xlabel}[][][0.70]{$r$ (m)}
\hspace{-0.8cm}~
\epsfig{file=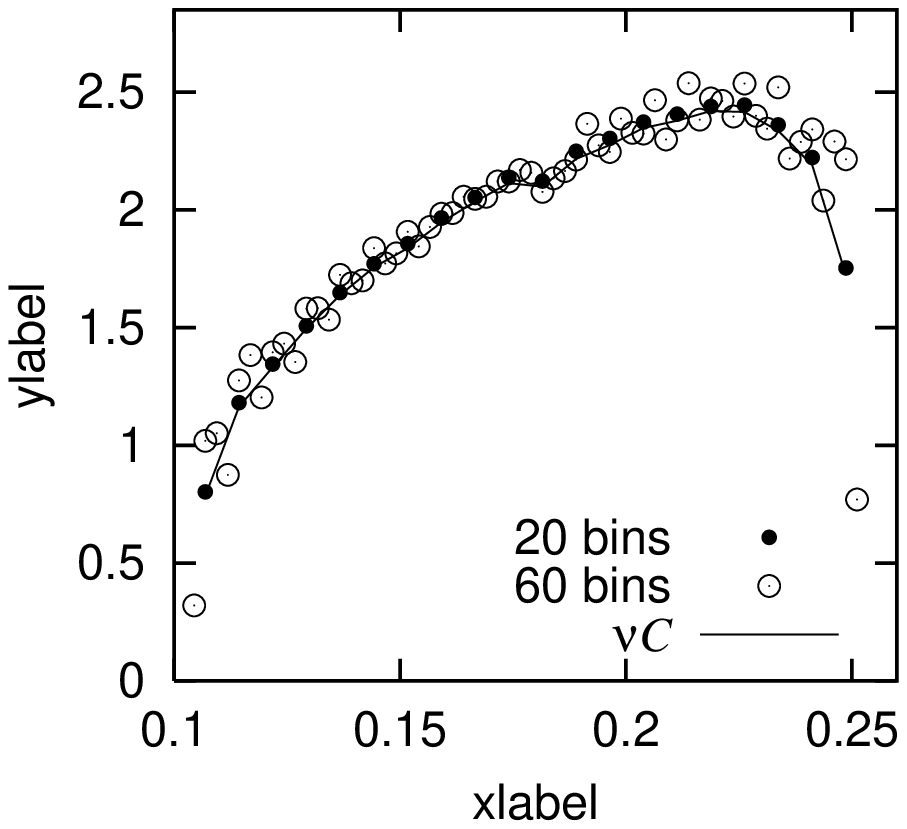,width=4.0cm,angle=-90}
\hspace{-0.4cm}~
\epsfig{file=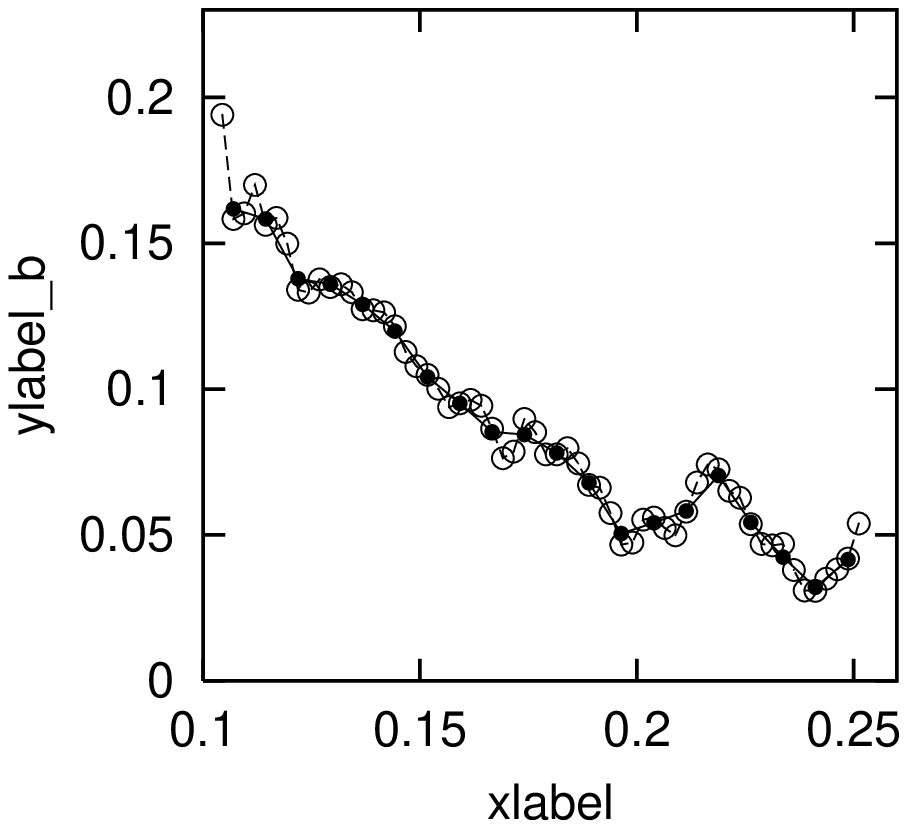,width=4.0cm,angle=-90}
~\hspace{-1cm}
~\vspace{0.1cm}\\
\caption{Components of the fabric tensor $\tr(\matrix{F})$ (isotropic)
and $\dev(\matrix{F})/\tr(\matrix{F})$ (anisotropic) for different
numbers of binning intervals
}
\label{fig:fabric}
\end{figure}

The trace of the fabric tensor is a measure for the contact number
density, whereas the deviator of the fabric is a measure for the 
anisotropy of the contact network \shortcite{latzel00,latzel00b}, 
see Fig.\ \ref{fig:fabric}.
In our situation, the contact number density is reduced in the shear
zone and the anisotropy, i.\,e.~the deviatoric fraction, is increased
in the shear zone, but remains below 20 per-cent.\\

\subsection{\em Stress Tensor}

\begin{figure}[htb]
\begin{center}
    \psfrag{ylabel}[][][0.95]{$\bsigma$ (N\,m$^{-2}$)}
    \psfrag{xlabel}[][][0.95]{$r$ (m)}
\epsfig{file=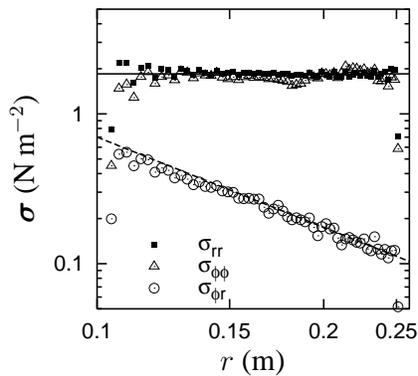,width=5.2cm,angle=-90}
\end{center}
\caption{Components of the static stress $\bsigma$
plotted against the distance from the center $r$ in log-log representation.
}
\label{fig:stress}
\end{figure}

In Fig.\ \ref{fig:stress}, the static stress components
are plotted. Here, the diagonal elements are almost constant, 
whereas the off-diagonal elements decay proportional to $r^{-2}$, 
as indicated by the lines, in consistency with continuum theoretical
equilibrium conditions \shortcite{luding00d}.  
The trace of the stress tensor, i.\,e.\ the volumetric stress, is almost
constant over the whole shear-cell besides fluctuations. In contrast,
the deviatoric fraction $\sigma_D/\sigma_V$ decays with increasing 
distance $r$ from the inner ring, similar to the behavior of
the deviatoric fraction of $\matrix{F}$.

\subsection{\em Tensor Orientations}
\label{sec:material_properties}

The orientations of the tensors  $\matrix{F}$, $\bsigma$, and 
$\bvarepsilon$ (data not shown here) are tilted from the radial outwards 
direction into shear direction 
\shortcite{latzel00,luding00d,latzel00b}.  
Interestingly, the orientations of the 
three tensors are different, indicating an anisotropic material.
All orientation angles show the same qualitative
behavior, however, the fabric is tilted more than the stress which, in
turn, is tilted more than the deformation gradient. Thus, the three
tensorial quantities examined so far are \emph{not} co-linear.


\subsection{\em Angular velocity field}

\begin{figure}[htb]
  \begin{center}
    \psfrag{xlabel}[][]{$\tilde{r}$}
    \psfrag{ylabel}[][]{}
    \psfrag{-Wrp}[][][0.8]{$-\omega_c$}
    \psfrag{-w}[][][0.8]{$-\omega$}
    \psfrag{-w*}[][][0.8][-90]{$-\omega^*$}
\psfrag{3358}[][][0.65]{0.8083}
\psfrag{3386}[][][0.65]{0.8149}
\psfrag{3401}[][][0.65]{0.8194}
    \epsfig{file=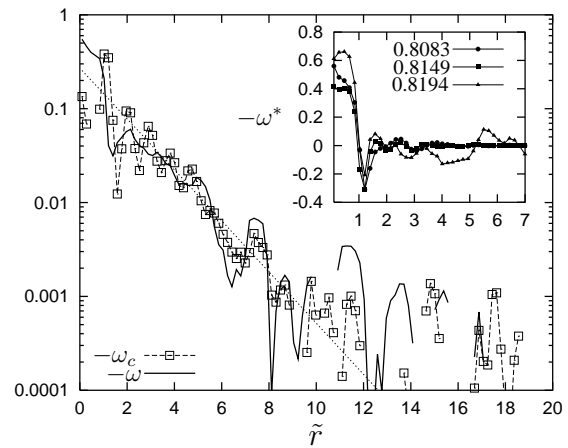,width=6cm,angle=-90,clip=true}
  \end{center}
  \caption{Angular velocities $\omega$ ({\it solid line}) and
    $\omega_c$ ({\it symbols}) of the particles and the continuum,
    plotted against the scaled radial
    distance (from simulation B). The {\it dotted} line is $\omega_c$
    as obtained from the fit to $v_{\phi}$, see
    subsection~\ref{sec:massflux}. In the inset, the excess-spin is
    displayed for all simulations}
  \label{fig:omega}
\end{figure}

In Fig.~\ref{fig:omega} the ($z$-components of the) macroscopic particle
rotations $\omega$, the continuum rotation $\omega_c$ and the 
particle excess-rotation $\omega^*=\omega - \omega_c$, are displayed. 
Both the particle- and the
continuum-angular velocity decay exponentially with increasing $\tilde{r}$,
paralleling the behavior of the velocity $v_{\phi}$. The inset of 
Fig.~\ref{fig:omega} shows an oscillation of the excess-rotation near the 
inner wheel, from one disk layer to the next. This is due to the fact that
the disks in adjacent layers are able to roll over each other in the shear zone.

\section{MATERIAL PARAMETERS}
\label{sec:material_parameters}

\subsection{\em Bulk Stiffness}

\begin{figure}[htb]
\begin{center}
  \psfrag{xlabel}[][]{tr$({\bf F})$}
  \psfrag{ylabel}[][]{$2\pi \bar{E}/k_n'$}
   ~\vspace{-1.0cm}\\
  \epsfig{file=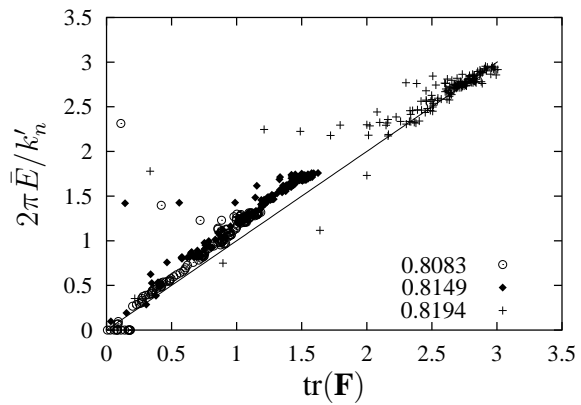,width=6cm,angle=-90,clip=true}
\end{center}
\caption{Scaled granular bulk stiffness $2 \pi \bar E/k_n'$, plotted 
  against $\tr(\matrix{F})$. Every point corresponds to one of 150
  rings dividing the system }
\label{fig:macro1}
\end{figure}

The material stiffness $\bar E = \tr(\bsigma)/\tr(\bvarepsilon)$
is here defined as the ratio of the volumetric parts of stress and 
elastic deformation gradient. In a mean field estimate it is found to
be proportional to the ``microscopic'' contact stiffness $k_n'$
and the trace of the fabric tensor:
$\bar E \propto (k_n' /2\pi) \, \tr(\matrix{F}).$
In Fig.~\ref{fig:macro1} the rescaled stiffness of the granular material
is plotted against the trace of the fabric.  Note
that all data collapse almost on a line, but the mean-field value
underestimates the simulation data by some per-cent.  
The few data points which deviate most are close to the boundaries. 
The deviation from the mean field prediction ({\it solid line} in
Fig.~\ref{fig:macro1}) seems to disappear in the absence of shear. 

\subsection{\em Shear Stiffness}

\begin{figure}[htb]
  \begin{center}
    \psfrag{xlabel}[][]{tr$({\bf F})$}
    \psfrag{ylabel}[][]{$\pi G/k_n'$}
     ~\vspace{-1.0cm}\\
    \epsfig{file=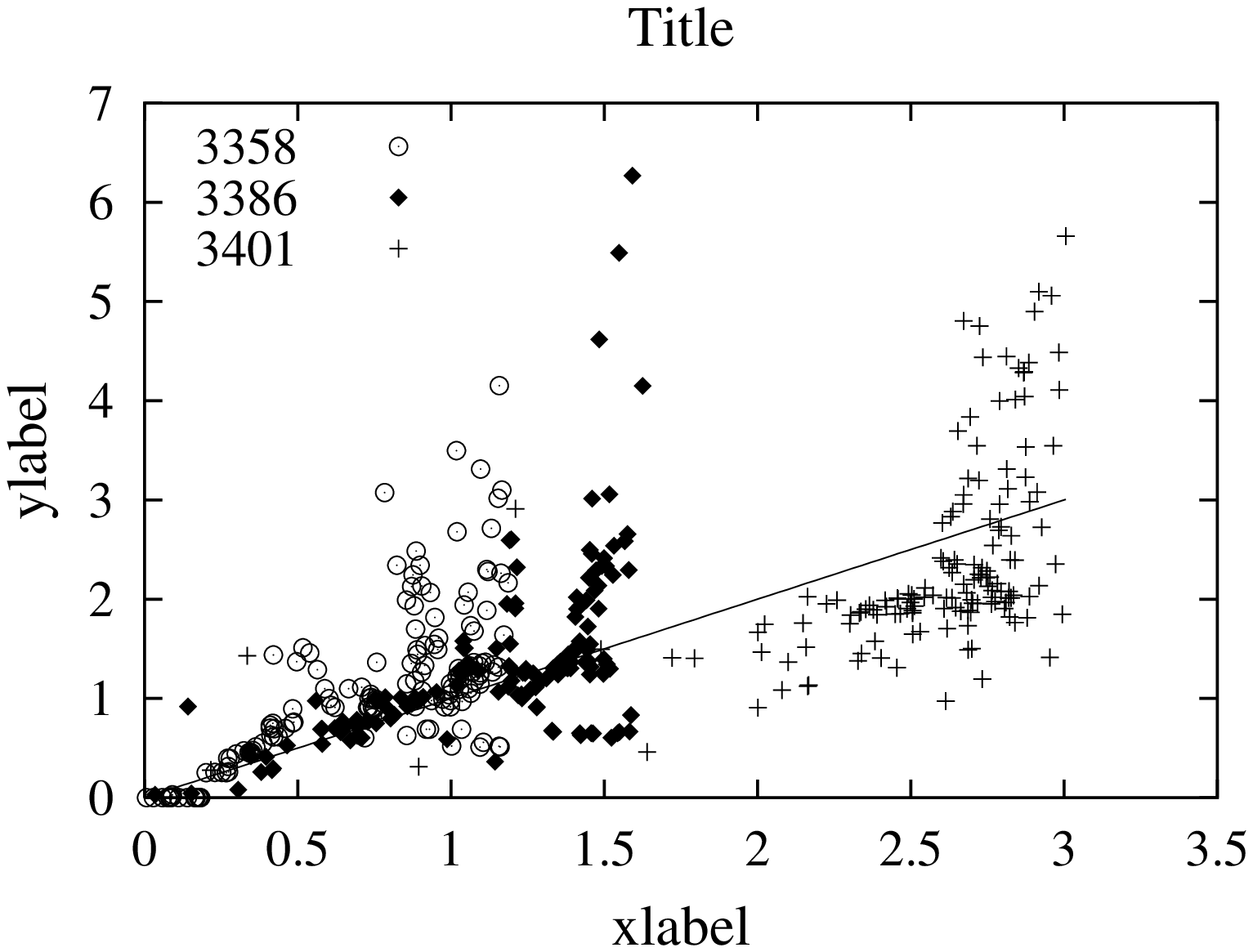,width=6cm,angle=-90,clip=true}
  \end{center}
  \caption{Scaled granular shear resistance 
    $\pi G/k_n'=\dev(\bsigma)/\dev(\bvarepsilon)$ plotted against
    $\tr(\matrix{F})$ from all simulations. The line indicates the
    identity curve }
  \label{fig:macro2}
\end{figure}

The typical shear stiffness of the material is defined as the
ratio of the deviatoric parts of stress and elastic deformation 
gradient $G = \dev(\bsigma)/\dev(\bvarepsilon)$ and scales in 
a crude mean field approximation as $G \propto (k_n'/\pi) \tr(\matrix{F})$.
In Fig.~\ref{fig:macro2} the ratio of the deviatoric parts of stress
and strain is plotted against the trace of the fabric.  
Like the bulk stiffness, both quantities are proportional, at least for points
near or within the dilute shear band. In the denser outer part of the 
shear-cell, the particles are strongly inter-locked and thus resist much 
more against shear, so that $G$ diverges. The critical contact number 
density grows with increasing global density, i.\,e.\ with increasing stress.

\subsection{\em Torque Resistance}

\begin{figure}
\begin{center}
  \psfrag{xlabel}[][]{$\tilde{r}$}
  \psfrag{ylabel}[][]{$\mu_c$}
   ~\vspace{-1.0cm}\\
  \epsfig{file=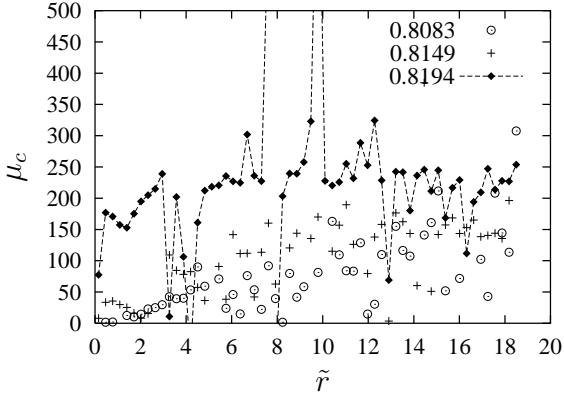,width=6cm,angle=-90,clip=true}
  \caption{Torque resistance plotted against $\tilde{r}$ }
  \label{fig:torque}
\end{center}
\end{figure}

Since we are interested in the role the rotational degree of freedom
plays for the constitutive equations, we define a new material parameter,
the ``torque resistance'' $\mu_c$, as the ratio of the magnitudes of the 
couple stress and the curvature components. This quantity describes how
strongly the material resists against applied torques. In
Fig.~\ref{fig:torque}, $\mu_c$ is plotted against the scaled
distance from the inner ring.
In the dilute regions near to the shear zone, where the
particles are able to rotate more easily, $\mu_c$ is smaller than in
the denser regions, where the particles are interlocked and thus
frustrated. This is consistent with the results for
increasing global densities, i.\,e.\ the torque resistance increases with
density. Note that the strongest fluctuations are due to the division by
small values and have no physical relevance in our interpretation.

\section{SUMMARY AND CONCLUSION}

A micro-macro averaging procedure was presented and specific
boundary conditions were chosen to allow for averaging over large
volumes and long times. A shear band is obtained, localized close to the
inner, rotating cylindrical wall. The local configurations changed rather
rapidly in the shear band, whereas the system is almost frozen in the 
outer part.

The averaging strategy used assumes the quantities to be homogeneously
smeared out over the whole particle which is cut in slices by the
averaging volumes.  
The material density, i.\,e.\ the volume fraction, the velocity field,
the fabric tensor, the stress tensor and the elastic, reversible 
deformation gradient were obtained from DEM simulations.
The volumetric stress is constant in radial direction while its deviatoric 
fraction and also the mean velocity gradient decay with increasing 
distance from the inner wall. The
ratio of the volumetric parts of stress and strain gives the material
stiffness of the granulate, which is small in the shear band, due to
dilation, and larger outside.

In the shear-cell, large deviators, i.\,e.\ anisotropy, of all tensorial
quantities is evidenced. All tensors are tilted from the radial direction.
The system organizes itself such that more contacts are
created to act against the shear and also the shear resistance
increases with the contact density.  An essential result is that the
macroscopic tensors are {\em not} co-linear, i.\,e.\ their orientations
are different.  Thus, the material cannot
be described by a simple elastic model involving only the two Lam\'e
constants (or bulk modulus and Poisson's ratio). 

Finally, the particles' angular velocity was averaged in analogy to the
particle velocity. Subtraction of the continuum rotation from the
particle rotation leads to the excess-eigen rotation of the particles
with respect to the mean rotation, in the spirit of a micro-polar or
Cosserat continuum theory. In analogy to the stress and elastic
deformation gradient, we defined couple stress and curvature. The
quotient of the respective non-zero components gives a
``torque-resistance'' which increases with increasing local density
and stress.

Current research concerns the application of the averaging formalism
to other boundary conditions, the use of more realistic particle 
interaction models, and the measurement of other macroscopic quantities
not discussed here.

\section{ACKNOWLEDGEMENTS}

The author thanks M. L\"atzel for data and discussions and
acknowledges financial support by the Deutsche Forschungsgemeinschaft
(DFG).


\end{document}